\documentclass[twocolumn,showpacs,preprintnumbers,amssymb,prl]{revtex4}
\usepackage{graphicx}
\usepackage{dcolumn}
\usepackage{color}
\usepackage{bm}

\def \be{\begin{equation}}
\def \ee{\end{equation}}
\def \ben{\begin{eqnarray}}
\def \een{\end{eqnarray}}

\begin{document}

\title{Estimating Temperature Fluctuations in the Early Universe}

\author{Debashis Gangopadhyay}
\altaffiliation{debashis@bose.res.in}
\affiliation{S. N. Bose National Centre for Basic Sciences,
         JD-Block, Sector-III, Salt Lake, Kolkata 700098,India\\
         and Centre for Astroparticle Physics and Space Science,\\
         Bose Institute (Salt Lake Campus), EN-Block, Salt Lake, Kolkata-700091, India}

\begin{abstract}
A lagrangian for the $k-$ essence field is constructed for a constant scalar potential
and its form determined when the scale factor was very small compared to the present epoch
but very large compared to the inflationary epoch. This means that one is already in an
expanding and flat universe. The form is similar to that of an oscillator with
time-dependent frequency. Expansion is naturally built into the theory with the existence
of growing classical solutions of the scale factor. The formalism allows one to estimate
fluctuations of the temperature of the background radiation in these early stages
(compared to the present epoch) of the universe. If the temperature at time $t_{a}$ is $T_{a}$
and at time $t_{b}$ the temperature is $T_{b}$ ($t_{b}>t_{a}$), then for small times , the probability for
the logarithm of  inverse temperature evolution can be estimated to be given by
$$P(b,a)=
|\langle ln~({1\over T_{b}}),t_{b}| ln~({1\over T_{a}}),t_{a}\rangle|^{2}$$
$$\approx\biggl({3m_{\mathrm Pl}^{2}\over \pi^{2} (t_{b}-t_{a})^{3}}\biggr)
(ln~ T_{a})^{2}(ln~T_{b})^{2}\biggl(1 - 3\gamma (t_{a} + t_{b})\biggr)$$
where  $0<\gamma<1$, $m_{\mathrm Pl}$ is the Planck mass and Planck's constant and the
speed of light has been put equal to unity.
There is the further possibility that a single scalar
field may suffice for an inflationary scenario as well as the dark matter and dark energy realms.


\end{abstract}


\maketitle

{\bf 1.Introduction}

An  approach to understand the origins of dark matter and dark energy involve
setting up lagrangians for what are known as $k-$essence fields. 
Usually one takes the Friedman-Robertson-Walker metric with zero curvature constant
so that the universe is flat. Recently a lagrangian for the  $k-$ essence 
field has been set up \cite{gangopadhyay} where 
there are two generalised coordinates $q(t)= ln~a(t)$ ($a(t)$ is the scale factor) 
and a scalar field $\phi(t)$ with a complicated polynomial interaction between them. 
In the lagrangian, $q$ has a standard kinetic term while $\phi$ does not have a
kinetic part and occurs purely through the interaction term.
Classical solutions of this lagrangian give very good results for the
cosmological parameters \cite{gangopadhyay}. It had been shown previously
\cite{scherrer} that it is possible to unify the dark matter and dark energy
components into a single scalar field model with the scalar field $\phi$ having a
non-canonical kinetic term. These scalar fields are the $k-$essence fields
which first appeared in models of inflation \cite{armendariz} and subsequently
led to models of dark energy also \cite{chiba}.
The general form of the lagrangian for these $k-$essence models is assumed to
be a function $F(X)$ with $X=\nabla_{\mu}\phi\nabla^{\mu}\phi$, and do not depend explicitly
on $\phi$ to start with. In \cite{scherrer}, $X$ was shown to satisfy a general
scaling relation,{\it viz.} $X ({dF\over dX})^{2}=Ca(t)^{-6}$ with $C$ a
constant (similar expression was also derived in \cite{chimento1}).
Using this, in \cite{scherrer}
specific forms (motivated from string theory \cite{gasperini},
\cite{armendariz}, \cite{sen}) were assumed for the lagrangian and $F(X)$ to show that
self-consistent models can be built which account for both the dark matter and
dark energy components. \cite{gangopadhyay} incorporates the scaling relation
of \cite{scherrer}. Literature on dark matter, dark energy and $k-$ essence
can be found in \cite{sahni},\cite{peebles},\cite{malquarti}, \cite{chimento2},\cite{abramo}.

The motivation for this work comes from the fact that as classical solutions of the
lagrangian in \cite{gangopadhyay} give realistic results , addressing  questions
relating to (quantum) fluctuations becomes meaningful. In fact, questions regarding the
amplitude of a scale factor at some epoch evolving to a different value at a later
epoch can surely be addressed for  a lagrangian with canonical kinetic terms. 
These effects are relevant at times close to the big bang (very small scale factor). 
However, then the curvature constant is not zero. To get past this difficulty,
in this work, an epoch is chosen  when the scale factor was small compared 
to the present epoch but very large compared to the big bang and inflationary epoch. 
This means that one is already in an expanding and flat universe-- however,the 
influence of inflation is still around. 

In this work, the scalar potential $V(\phi)=V$ is a ({\it positive}) constant and all the 
time variables $t\equiv t/t_{0}$, where $t_{0}$ is the present epoch and we are 
interested only in $t< 1$ scenarios. Also $a(t_{1})< a(t_{2})$ for $t_{1}< t_{2}$ etc.
So one has to find how the lagrangian looks at smaller values of the scale
factor,find possible classical solutions  and then find relevant fluctuations.
As the scale factor is inversely proportional to the  temperature at a particular
epoch, these amplitudes also provide an estimate of temperature  fluctuations.
In this work , I discuss a plausible formalism for this and show that the classical 
solutions in this domain are fully consistent with the inflationary scenario and a 
radiation dominated universe.Moreover, a single scalar field may suffice for an 
inflationary scenario as well as the dark matter and dark energy realms 
and a measure of temperature fluctuations may be estimated using standard prescriptions.

{\bf 2. Lagrangian in the early universe}
\label{}

The lagrangian $L$ (or the pressure $p$) is taken as
$$ L= -V(\phi) F(X)\eqno(1)$$
The energy density is
$$\rho = V(\phi)[ F(X) -2 X F_{X}]\eqno(2)$$
with $F_{X}\equiv {dF\over dX}$ and in the present work $V(\phi)=V$ is a constant ($> 0$).
Before proceeding further, a digression on the origin of 
the equation $(1)$ is in order. The present view is that dark energy can account for the 
observed acceleration of the galaxies. One can show that dark energy with sufficiently 
negative pressure will accelerate the expansion of the universe once it starts dominating 
over matter. A simple fluid model with negative presure is the cosmological constant 
with equation of state parameter $w=-1$, and energy density $\rho = -p $ ,the pressure. 
If one identifies the dark energy with a cosmological constant then a fundamental length 
is introduced into the theory \cite{sahni} 
$L_{\Lambda}\equiv H_{\Lambda}^{-1}$ related to the dark energy density 
$\rho_{DE}$ by $H_{\mathrm \Lambda}^{2}\equiv {8\pi G\rho_{DE}\over 3}$. Observations 
then require that 
$(L_{\mathrm Pl}^{2}/ L_{\mathrm\Lambda}^{2})\leq 10^{-123}$, 
where $L_{\mathrm Pl}$ 
is the Planck length. So there is an enormous fine tuning problem. There are also other 
problems. Alternative possibilities entail use of scalar fields $\phi$ with suitably chosen 
potential  $V(\phi)$ so as to have variable (with time) vacuum energy. The hope is then that 
fine tunings can be avoided with suitably chosen potentials. Two such models with variable 
$w$ are 
$ L_{\mathrm quintessence}= {1\over 2}\partial_{\mu}\phi\partial^{\mu}\phi - V(\phi)$ and
$ L_{\mathrm tachyon}= -V(\phi)[1-\partial_{\mu}\phi\partial^{\mu}\phi]^{1\over 2}$.
The quintessence model can be shown to be a natural generalisation of the lagrangian for a 
non-relativistic particle \cite{sahni}. In a cosmological context , the 
quintessence (scalar field) models leads to potential energy driven acceleration 
($V(\phi) > \partial^{\mu}\phi\partial_{\mu}\phi)$.
The tachyon lagrangian originated from string
theory \cite{sen} and can be shown to be a field theoretic analogue of the lagrangian 
of a relativistic particle \cite{sahni}. The model allows solutions 
where $V\rightarrow 0, \partial_{i}\phi\partial^{i}\phi\rightarrow 1$ simultaneously, keeping 
the energy density finite. So here the kinetic energy dominates and in the cosmological 
context can give rise to acceleration. These are the {\it k-essence} models for dark energy.
In the cosmological scenario, the $[1-\partial_{\mu}\phi\partial^{\mu}\phi]^{1\over 2}$ factor 
is replaced by a more general function $F(X)$ with $X=\partial^{\mu}\phi\partial_{\mu}\phi$.
This is the genesis of equation $(1)$.

For a flat Robertson Walker metric the equation for the $k-$essence field is
$$(F_{X}+2XF_{XX})\ddot\phi+3HF_{X}\dot\phi+(2XF_{X}-F){V_{\phi}\over V}=0\eqno(3a)$$
$H={\dot a(t)\over a(t)}$ is the Hubble parameter.
For a homogeneous and isotropic universe $\phi(x,t)\equiv\phi(t)$, and so 
$X={1\over 2} \dot\phi^{2}$. In \cite{scherrer}, $V(\phi)=const.$ so $V_{\phi}\equiv {dV\over d\phi}=0$
and $(3a)$ becomes
$$(F_{X}+2XF_{XX})\dot X +6HF_{X}X=0\eqno(3b) $$
Changing the independent variable from the time $t$ to the
scale factor $a$ i.e. ${dX\over dt} = a H {dX\over da}$ leads to \cite{scherrer}
$$(F_{X}+2XF_{XX})a{dX\over da} +6F_{X}X=0\eqno(3c) $$
whose solution is the scaling law \cite{scherrer} :
$$X F_{X}^{2}=Ca^{-6}\eqno(4)$$
In this work also we take $V(\phi)=V= constant$ so that the above scaling law holds. 

Using $(4)$ and the zero-zero component of Einstein's field equations an expression 
for the lagrangian is obtained as follows.Take the Robertson-Walker metric :
$$ds^{2}= c^{2}dt^{2} - a^{2}(t)[{dr^{2}\over (1-kr^{2})} + r^{2}(d\theta^{2} 
+ sin^{2}\theta d\phi^{2})]\eqno(5)$$
where $k=0, 1\enskip or -1$ is the curvature constant.
The zero-zero component of Einstein's equation reads:
$$R_{00} - {1\over 2}g_{00}R = - \kappa T_{00}\eqno(6)$$
This gives with the metric $(5)$
$${k\over a^{2}} + H^{2} = {8\pi G\over 3}\rho\eqno(7)$$
Using $(1),(3),(6)$ and $(7)$,and taking $k=0$, we arrive at
$$X F_{X}= {1\over 2X}[ F - ({3\over 8\pi G V)})H^{2}]\eqno(8)$$
Using $(4)$ to eliminate $F_{X}$  gives
$$F(X) = 2{\sqrt C}{\sqrt X}a^{-3} + 3 {H^{2}\over 8\pi G V}\eqno(9)$$
So the  expression for the lagrangian is obtained as
$$L=-2{\sqrt C}{\sqrt X}a^{-3}V - ({3\over 8\pi G}) H^{2}$$
$$=-2{\sqrt C}\sqrt {\dot\phi ^{2}-(\nabla\phi)^{2}}a^{-3}V
- 3{H^{2}\over 8\pi G}\eqno(10)$$
Homogeneity and isotropy
of spacetime imply $\phi(t,\bf x)=\phi(t)$.
Then $(10)$ becomes
$$L=-c_{1}\dot q^{2} - c_{2} V\dot \phi e^{-3q}\eqno(11)$$
where $a(t)=e^{q(t)}$, $c_{1}= 3(8\pi G)^{-1}$, $c_{2}=2 \sqrt C$,
(we shall always take the positive square root of$C$) 
and the scalar potential $V$ is a constant. 

The lagrangian $(11)$ has a kinetic term for $q$. The other term 
is an interaction term and there is no kinetic term for $\phi$. 
Note that $a=e^{q}$. It is readily seen from the graph of the exponential function 
(in figure 1 $x\equiv q$) that in the region  $-1 < q < 0$ one has , $a=e^{q} < 1$ . Hence in this region 
$q$ small (i.e. $\vert q\vert <1$) means $a$ is also small (i.e $a < 1$). Moreover, in this region 
$a$ grows from $e^{-1}=0.367879$ to $e^{0}=1$. So within this region $a$ 
grows as $q$ grows. So smaller values of $q$ mean that we are going back to 
smaller values of $a$ i.e. to earlier epochs. In this work we will restrict ourselves to this
domain. In this domain 
expand the exponential in $(11)$ :
$$L=-c_{1}\dot q^{2} 
-[{2\sqrt C}V\dot\phi][1 - 3q + {9\over 2}q^{2} +...]\eqno(12)$$ 
Keeping terms upto $O(q^{2})$ and replacing $q$ by
$q+ {1\over 3}$ one has
$$L = -{M\over 2}[\dot q ^{2} + 12\pi G g(t) q^{2}] -({1\over 2})g(t)\eqno(13)$$
where $M={3\over 4\pi G} = {3m_{\mathrm Pl}^{2}\over 4\pi}$,
$g(t)= 2{\sqrt C}V \dot\phi$, $m_{\mathrm Pl}$ is the Planck energy
and we use $\hbar = c = 1$ ($c$ is speed of light,$\hbar$ is Planck's constant).

Here two clarifications must be made. Firstly, $q= ln a$ means $\dot q = {\dot a\over a}= H(t)$. 
In the present work {\it the Hubble parameter $H(t)$ is not a constant but a function of 
time} as it usually is.

Secondly,although in $(12)$ the lagrangian is an explicit 
function of $a,\dot a$ , it would be incorrect to assume that one should have this 
imposed right from equation $(1)$ itself. 
This is because the dependence on $a,\dot a$ 
follows {\it only after} the following operations have been carried out , {\it viz.},

(a) the Robertson-Walker metric is chosen and the equation for the $k-$ essence field 
is set up for a constant $V(\phi)$ ; 

(b)a change of the independent variable is made from the time $t$ to the scale factor $a$; 

(c) the scaling law $(4)$ obtained; 

(d) the zero-zero component of Einstein's equation incorporated to relate the energy density 
to the curvature constant $k$ (subsequently we take $k=0$) and the Hubble parameter; and 

(e) the scaling law $(4)$ used to eliminate the derivative $F_{X}$. 

Therefore, one must start with an independent potential in the lagrangian and let the 
relevant equations dictate ({\it a posteriori}) the final form for the lagrangian. 
In this method there is no possibility of non-Einsteinian gravity equations emerging. 
It is shown below (equations $(21), (22)$) that the the classical solution for the 
scalar field is linear in the time $t$ so that in $L_{\mathrm cl}$ the term proportional 
to $\dot\phi_{\mathrm cl}$ is effectively a constant without any dynamical consequences.
In fact, all total derivatives will be ignored (see below).

{\bf 3.Chaotic Inflation}

First consider a possible classical solution in this model.
Put  $12\pi G g(t) = -\Omega^{2}(t)$.
This means
$$\phi(t)= -{1\over 24\pi G {\sqrt C}V}\int dt \Omega ^{2}(t)\eqno(14)$$
$(13)$ now becomes
$$L =-{M\over 2}[\dot q ^{2} - \Omega^{2} (t) q^{2}] -({1\over 2})g(t)\eqno(15)$$
Now, the term ${1\over 2}g(t)= {\sqrt C}V{d\phi\over dt}$
is a total time derivative
and thus has no contribution to the equations of motion and hence ignorable.
Then$(15)$ becomes
$$L = -{M\over 2}[\dot q ^{2} 
- \Omega^{2} (t) q^{2}]\eqno(16a)$$
So the problem of evolution of the scale factor
in a homogeneous universe at early times (i.e. early compared to our present epoch but late 
compared to the big bang) may be studied by considering an
oscillator with time-dependent frequency. 

Let us recollect what we have done. We started with a lagrangian where there was 
a dynamical degree of freedom $q$ and a scalar field without a kinetic term. 
We then examine the lagrangian in a region where $q$ is small so that terms 
upto $O(q^{2})$ are only kept and we have an oscillator.Interestingly, the role 
of the scalar field is transcribed into a time dependent frequency for our oscillator! 

Thus, occurrence of parametric resonance
becomes a distinct possibility.Parametric resonance in the early universe has been
reported before in a different context. A theory of preheating and reheating after inflation
was developed based on parametric resonance in the context of a scalar field $\chi$ coupled
to an inflaton field \cite{kofman}.

It should be noted that the effect of the scalar field $\phi$ has been re-written
as a time dependent frequency $\Omega(t)$,
($\Omega^{2}(t)=-24\pi G{\sqrt{C}}V{d\phi\over dt}$). With this
the lagrangian now looks like that of a time dependent oscillator and
one of the  solutions to this system is that of the parametric oscillator.
First let us assume that the time dependent frequency is obtained as a
relatively small sinusoidal perturbation
from that of the ideal oscillator i.e.
$$\Omega^{2} (t)=-24\pi G{\sqrt{C}}V{d\phi\over dt}\equiv \omega_{0}^{2} [1+f(t)]$$
where   $f(t)= h~cos\gamma t$ , with the constant $h << 1$ and positive 
and $\omega_{0}^{2}> 0 $ ,i.e. $\omega_{0}$ real. Note that 
we will always be in the realm of small postive $t$ ($t < 1$) and therefore 
$$\Omega^{2}(t)=-24\pi G{\sqrt{C}}V{d\phi\over dt}=\omega_{0}^{2}+\omega_{0}^{2}h~cos\gamma t\eqno(16b)$$
and $\Omega ^{2}(t)$ will always be positive. (This means that for consistency $\dot\phi$ has to be 
negative and this consistency is present in equation $(17)$ below.)  

The physical motivation for taking $f(t)$ to be sinusoidal is as follows. 
A sinusoidal $f(t)$ can give growing solutions for $q$ and since $a=e^{q}$ we also  
get growing solutions for the scale factor which we know holds for an expanding universe. So the 
scalar field also helps in an inflation-like scenario. Note that the same scalar field can 
account for (a)growing solutions for $a$ (shown below , equation $(20)$) 
(b)dark matter and dark energy as shown 
in \cite{gangopadhyay} and (c)a simple scheme for estimating temperature fluctuations in certain 
epochs (Section 4). So the goal of building realistic models with a minimum number of scalar 
fields is also possible with this choice for $f(t)$. 		 

Putting this $f(t)$ in equation $(14)$ gives for small values of time $t$
$$\phi (t)\approx\phi_{ini} - {\omega_{0}^{2} (1+h)\over 24\pi G{\sqrt C}V} t\eqno(17)$$
where $\phi_{ini}$ is a constant of integration and may be identified with 
the initial value of $\phi$.

The excitation (forcing) term has amplitude $h$ and period
$T_{exc}= {2\pi\over 2\omega}={\pi\over\omega}$. The unexcited
(natural)period is $T_{nat}= {2\pi\over \omega_{0}}$. Now solutions of
linear ordinary differential equations with periodic coefficients \cite{landau}
have following properties:
(a) if the coefficients are periodic with period T , then if $q(t)$ is a solution,so is $q(t+T)$\\
(b) Any solution $q(t)= A q_{1}(t)+ B q_{2}(t)$ where $q_{1}(t), q_{2}(t)$ are two linearly 
independent solutions.$A,B$ come from initial conditions. 
Periodicity means that $q_{1}(t+T),q_{2}(t+T)$ are 
also solutions and it follows that $q(t+T)= \lambda q(t)$. So $q(t)$ is periodic within a scaling 
parameter $\lambda$. Consider only growing or decaying solutions ,{\it not oscillations}.
Then $q(t+T)=e^{\mu T}q(t)$ where $\mu = {ln|\lambda|\over T}$.
Finally, define $P(t)$ to be periodic such that $P(t+T)=P(t)$. Then
$$q(t) = e^{\mu t}P(t)\eqno(18)$$
is the general form of the solution with the stability of the solution depending 
on the sign of $\mu$. Seek a solution of the general form
$q(t)= A(t)cos (\omega_{0} + {1\over 2}\epsilon)t + B(t)sin (\omega_{0} + {1\over 2}\epsilon)t$.
For $f(t)$ as given above and $\gamma=2\omega_{0} +\epsilon, \epsilon << \omega_{0}$,
a solution of the general form \cite{landau} is
$$q(t)\equiv e^{\mu t}P(t)
= A_{0}e^{\mu t}cos (\omega_{0} + {1\over 2}\epsilon)t$$
$$+ B_{0}e^{\mu t} sin(\omega_{0} + {1\over 2}\epsilon)t\eqno(19)$$
where $\mu^{2}= {1\over 4}[({1\over 2}h\omega_{0})^{2}-\epsilon^{2}]$ and
$A_{0},B_{0}$ have to be determined from initial conditions.
Here we have retained terms that are
linear in $\epsilon$ and first order in $h$. Parametric resonance occurs in the range
$-{1\over 2}h\omega_{0} < \epsilon < {1\over 2}h\omega_{0}$ on either side of $2\omega_{0}$.
The region of instability as well as the amplification coefficient $\mu$ are of the order of $h$.
Parametric resonance also occurs when the frequency $\gamma$ is close to any value
${2\omega_{0}\over n}$ with $n$ integral. The region of instability and the amplification
decreases rapidly with $n$.

For very small time scales the classical solutions for the scale factor
of $(16)$ can be written as (upto $O(t^{2})$ terms):
$$a_{c}(t) = e^{q_{c}(t)}\approx a_{i} e^{At -  Bt^{2}}
\eqno(20)$$
where $a_{i}=e^{A_{0}}$ is the initial value for $t=t_{i}\sim 0$,
$A=\mu A_{0} +\omega B_{0}$ , $B= {1\over 2} A_{0}(\omega^{2}-{\mu}^{2}) - B_{0}\mu\omega$ and
$\omega\equiv \omega_{0} + {1\over 2}\epsilon$. So we have growing solutions for $a$ and 
{\it not oscillating solutions}. So there is no question of expanding and contracting scale 
factors. This solution is consistent with inflation.
Taking the initial time as $t_{i}<< 1$ and
any observable epoch $t_{0}\sim 1$,  the condition for 
an inflation like scenario,
{\it viz.}, ${\dot a_{i}\over\dot a_{0}}< 1$ is readily seen to be satisfied.
The expansion here is similar to "chaotic
inflation" \cite{linde}  where the initial
conditions of inflaton are distributed chaotically. For an inflaton potential
$V(\phi)={1\over 2} m^{2}\phi^{2}$, the inflaton field $\phi$ and the
scale factor are given by \cite{shinji}
$$\phi\approx \phi_{i} - {m m_{\mathrm Pl}\over 2{\sqrt { 3\pi}}}t$$
$$a\approx a_{i} exp \biggl[ 2 {\sqrt {\pi\over 3}}{m\over m_{\mathrm Pl}}
\biggl( \phi_{i} t - {m m_{\mathrm Pl}\over 4{\sqrt {3\pi}}}t^{2}\biggr)\biggr]
\eqno(21)$$
Comparing with $(21)$ , equations $(17)$ and $(20)$
will mimic the scenario of "chaotic inflation" with the following identifications:
$$
\phi_{i}=\phi_{ini}~ ;~
{{\omega_{0}^{2}(1+h)\over {\sqrt C}V}}= 2{\sqrt {3\pi}}({m\over m_{\mathrm Pl}})$$
$$A=\mu A_{0} + (\omega_{0} + {1\over 2}\epsilon) B_{0} = 2 {\sqrt {\pi\over 3}}({m\over m_{\mathrm Pl}}) \phi_{i}$$
$$B={A_{0}[(\omega_{0} + {1\over 2}\epsilon)^{2}-\mu^{2}]\over 2} - B_{0}\mu (\omega_{0} + {1\over 2}\epsilon) = {1\over 6} m^{2}
\eqno(22)$$
The solution can also be made consistent with a radiation dominated era as follows.
Using $(20)$,the Hubble parameter is $H = {\dot a_{c}\over a_{c}}= A-2Bt$,
so that the energy density is
$$\rho = {3 H^{2}\over 8\pi G}\approx ({3m_{\mathrm Pl}^{1\over 2}\over 8\pi })( A^{2}- 4AB t + 4B^{2}t^{2})
\eqno(23)$$
For a radiation dominated era  $\rho\sim a_{c}^{-4}$ with
$$a_{c}^{-4}\approx a_{i}^{-4} -4a_{i}^{-4} A t + a_{i}^{-4}(8A^{2}+4B)t^{2}
\eqno(24)$$
Equating the coefficients of the $O(t)$ terms in $(23)$ and $(24)$
gives $A^{2}=B$ and this combined with the second and third equations of $(22)$
imply the following for a radiation dominated
era {\it viz.}  $\phi_ {i} \sim {m_{\mathrm Pl}\over 8\pi} $.
This is consistent with "chaotic inflation" \cite{shinji}
where from considerations of the number of e-foldings required the 
bound on the initial value of the inflaton is obtained as $\phi_{ini}\geq 3m_{\mathrm Pl}$.
Equating the coefficients of $O(t^{2})$ terms in $(23)$ and $(24)$
and using $A^{2}=B$ gives $B={8\pi\over m_{\mathrm Pl}^{2}} e^{-4A_{0}}$.
This combined with the last equation of $(22)$ gives $A_{0}$.
$B_{0}$ is then obtained from the second equation of $(22)$ and
$A={\sqrt B}={m\over {\sqrt 6}}$.
(A consistency condition between $\mu$ and the last equation of $(22)$ 
also exists as $\mu ^{2}=(1/4)[(1/2)(h\omega_{0}^{2}-\epsilon^{2}]$).
Thus assuming
that $t_{i}<<1\sim 0$, one has
$$
A_{0}=q(t=t_{i})=ln~({2 (3\pi)^{1\over 4}\over (mm_{\mathrm Pl})^{1\over 2}}= ln~a(t_{i})$$
$$B_{0}={q(t={\pi\over 2\omega})\over e^{\mu\pi/2\omega}}
={ ln~a({\pi\over 2\omega})\over e^{\mu\pi/2\omega}}
= {m\over {\sqrt 6}\omega} -{\mu ln~a(t_{i})\over \omega}
\eqno(25)$$
Now $a(t=t_{i})\neq 0$
so that $q_{i}=ln~a_{i}$  is still well defined. At some later time  $t=t_{a}$ , let $q=q_{a}$.
Then the integration constants $A_{0}, B_{0}$ can be determined in terms of  $q_{i}$  and $q_{a}$.
Here $q_{i}\sim q(t_{i})$ and $q_{a}\sim q({\pi\over 2\omega})$.
The inflaton mass must be $m\approx 10^{-6}m_{\mathrm Pl}$ so as to fit the
observed amplitudes of density perturbations by COBE satellite \cite{lid}.
Therefore, the domain we are considering has the initial value of the scale factor as
$a(t_{i})\sim {2\times 10^{3}(3\pi)^{1\over 4}\over m_{\mathrm Pl}}$.

{\bf 4.Fluctuations}

Let us now discuss how to estimate (quantum) fluctuations in this model.
Given the structure of the lagrangian with canonical kinetic terms , 
this can always be done in principle.   
Write $q(t)=q_{\mathrm cl}(t) + x(t)$ where $x(t)$ is the fluctuation from the
classical value $q_{\mathrm cl}$ and $0\leq x(t)\leq\infty$. So
this corresponds to a time dependent oscillator on the half-plane \cite{khandlawa},
\cite{ezawa},\cite{simon},\cite{clark}. Then the quantum mechanical amplitude for
the (log of) scale factor to evolve from the value
$q_{a}$ at time $t_{a}$ to  $q_{b}$ at time $t_{b}$ is given by \cite{feyn}
$$
\langle q_{b},t_{b}| q_{a},t_{a}\rangle
=\langle ln a(t_{b}),t_{b}|ln a(t_{a}),t_{a}\rangle
= F(t_{b},t_{a})e^{{i\over\hbar}S_{cl}}
\eqno(26)$$
$S_{\mathrm cl}=\int ^{t_{b}} _{t_{a}} L_{\mathrm cl} dt
=\int_{t_{a}}^{t_{b}}dt [{M\over 2}\dot q_{\mathrm cl} ^{2} 
- {1\over 2}\Omega^{2}(t) q_{\mathrm cl}^{2}]$.
Here $\Omega(t)$ has been re-labelled as $\Omega(t)\rightarrow {\sqrt M}\Omega(t)$. 
(Factors of $\hbar$ and c will be put equal to unity at the end).
$F(t_{b}, t_{a})$ is calculated as follows \cite{khandlawa}.
The fluctuations $x(t)$ satisfy
$$\ddot x + \Omega^{2} (t) x = 0
\eqno(27)$$
As $\Omega$ is real, we will have quasi-periodic solutions. 

Here it may be mentioned that if one takes (say) the negative square root of $C$,
(or equivalently $\omega_{0}^{2} < 0$) then 
$\Omega^{2} < 0$ and we will have quasi-exponential solutions for $x$ i.e.
$x\approx A exp[-\int \vert\Omega(t)\vert dt] + B exp[+\int \vert\Omega (t)\vert dt]$. These solutions 
can be expected in the inflationary stage at which all second-order equations become elliptic. 
But in the present work all frquencies are real and we have only quasi-periodic solutions.  

Consider two independent quasi-periodic solutions of $(27)$:
$$x_{1}(t)=\psi (t)sin\zeta (t,t_{a})~;~ x_{2}(t)=\psi (t)sin\zeta (t_{b},t)
\eqno(28)$$
satisfying the boundary conditions
$$x_{1}(t_{a})\equiv x_{1a}=0~~;~x_{2}(t_{b})\equiv x_{2b}=0
\eqno(29)$$
where $\psi(t)$ satisfies the Ermakov-Pinney equation \cite{ermakov}
$$\ddot\psi + \Omega ^{2}(t)\psi - \psi ^{-3} =0
\eqno(30)$$
with $\zeta (t,s)$ defined as
$$\zeta(t,s) = \nu (t) - \nu (s) = \int ^{t}_{s} dt \psi ^{-2} (t)
\eqno(31)$$
$\psi(t)$ and $\nu (t)$ can be interpreted as the amplitude and phase of
the time dependent oscillator. Then the fluctation factor is given by
$$F(t_{b}, t_{a})
=\biggl({M\sqrt{(\dot\nu _{b}\dot\nu _{a})}\over 2\pi i\hbar sin\zeta (t_{b},t_{a})}\biggr)^{1\over 2}
\eqno(32)$$
Classical solutions satisfying relevant boundary conditions  are found ,
i.e. $q(t= t_{a})=q_{a}, q(t=t_{b})=q_{b}$ and the amplitude is
$$\langle q_{b},t_{b}| q_{a},t_{a}\rangle
=\langle ln~ a(t_{b})~,t_{b}|ln~ a(t_{a})~,t_{a}\rangle$$
$$=\biggl({M\sqrt{(\dot\nu _{b}\dot\nu _{a})}\over 2\pi i\hbar sin\zeta (t_{b},t_{a})}\biggr)^{1\over 2}
\biggl(exp({iS_{\mathrm cl}^{+}\over\hbar}) - exp({iS_{\mathrm cl}^{-}\over\hbar})\biggr)
\eqno(33)$$
where
$$S^{\pm}_{\mathrm cl}
=\biggl({\dot\psi_{b}q_{b}^{2}\over\psi_{b}}
-{\dot\psi_{a}q_{a}^{2}\over\psi_{a}}\biggr)$$
$$+{1\over sin\zeta (t_{b},t_{a})}
\biggl((\dot\nu_{b}q_{b}^{2}+\dot\nu_{a}q_{a}^{2}) cos\zeta (t_{b},t_{a})
\mp 2\sqrt{(\dot\nu_{b}\dot\nu_{a})}q_{b}q_{a}\biggr)
\eqno(34)$$
Now assume $\dot\nu <<1$
i.e. the time rate of change of phase is small.
Also note that the temperature of the background radiation in a
homogeneous universe is inversely proportional to the scale factor i.e.
$T(t_{a})\equiv T_{a}= {1\over a(t_{a})}$ (in appropriate dimensionless units).
Then to lowest orders of
$\dot\nu$, one has the probability for the logarithm of
scale factor ,or equivalently, logarithm of  inverse temperature evolution as
$$P(b,a)
=|\langle q_{b},t_{b}| q_{a},t_{a}\rangle|^{2}
=\biggl({3m_{\mathrm Pl}^{2}\over \pi^{2}\hbar^{4}c}\biggr)
\biggl({q_{a}^{2}q_{b}^{2}(\dot\nu _{b}\dot\nu _{a})^{3\over 2}\over sin\zeta ^{3}(t_{b},t_{a})}\biggr)$$
$$\equiv|\langle ln~({1\over T_{b}}),t_{b}| ln~({1\over T_{a}}),t_{a}\rangle|^{2}$$
$$=\biggl({3m_{\mathrm Pl}^{2}\over \pi^{2}}\biggr)
(ln~ T_{a})^{2}(ln~T_{b})^{2}
{(\dot\nu _{b}\dot\nu _{a})^{3\over 2}\over sin\zeta ^{3}(t_{b},t_{a})}
\eqno(35)$$
where $\hbar^{4}c$ has been put equal to unity.

Let us choose $\psi(t)=e^{\gamma t}$ where $0 <\gamma< 1$. Putting this in equation $(30)$, 
using $\Omega^{2}(t)=-24\pi G{\sqrt{C}}V{d\phi\over dt}$ and solving for $\phi$ gives 
$$\phi(t)={1\over 24\pi G{\sqrt{C}}V}[C_{0} + \gamma^{2}t + {1\over 4\gamma}e^{-4\gamma t}]$$
$$\approx (C_{0}+{1\over 96\gamma\pi G{\sqrt{C}}V})-{(1-\gamma^{2})\over 24\pi G{\sqrt{C}}V}t
\eqno(36)$$  
$C_{0}$ is a constant of integration.
Therefore the choice of $\psi(t)=e^{\gamma t}$ is consistent with what has been discussed 
before ({\it viz.} equations $(17)$, $(22)$) with the following identifications:
$$\phi_{\mathrm ini}\equiv C_{0}+ {1\over 96\gamma\pi G{\sqrt C}V}~~;~~   
\omega_{0}^{2}(1+h) =1 - \gamma^{2}\eqno(37)$$ 
Note that bounds on ${\sqrt C}V$ can be obtained in terms of $C_{0}, G,\gamma$ and
$\phi_{ini}\geq 3m_{\mathrm Pl}$ \cite{shinji} (discussion after equation $(24)$).
Further, with this choice for $\psi(t)$ , one has the following relations:
$$\nu(t)=- {1\over 2\gamma} e^{-2\gamma t}~~;~~ sin\zeta(t,s)=sin (\nu(t)-\nu(s))\approx t-s\eqno(38)$$
For small times , then the probability for the logarithm of  inverse temperature evolution,
$(35)$ becomes
$$P(b,a)
=|\langle ln~({1\over T_{b}}),t_{b}| ln~({1\over T_{a}}),t_{a}\rangle|^{2}$$
$$\approx\biggl({3m_{\mathrm Pl}^{2}\over \pi^{2}}\biggr)
{(ln~ T_{a})^{2}(ln~T_{b})^{2}\biggl(1 - 3\gamma (t_{a} + t_{b})\biggr)\over (t_{b}-t_{a})^{3}}
\eqno(39)$$
Let us estimate the physical domain of validity of this expression. By physical domain I mean 
the times $t_{a}, t_{b}$ and the temperatures $T_{a}, T_{b}$. As already stated in the introduction 
(section 1), all our times have to be understood as $t\equiv {t\over t_{0}}$.
The requirement that the $P(b,a)\geq 0$ and $0<\gamma<1$ means  
$$t_{a}+ t_{b}\geq {1\over 3}\eqno(40)$$  
Again the requirement $P(b,a)\leq 1$ implies 
$$t_{b}-t_{a}\leq \bigg({3m_{\mathrm Pl}\over\pi}[ln T_{a}][ln T_{b}]\biggr)^{2\over 3}\eqno(41)$$
Combining $(40)$ and $(41)$ gives 
$$ t_{a}\geq {1\over 6} -{1\over 2}\bigg({3m_{\mathrm Pl}\over\pi}[ln T_{a}][ln T_{b}]\biggr)^{2\over 3};$$ 
$$ t_{b}\geq {1\over 6} + {1\over 2}\bigg({3m_{\mathrm Pl}\over\pi}[ln T_{a}][ln T_{b}]\biggr)^{2\over 3}\eqno(42)$$ 
Now combining $(42)$ with the requirement that all estimates are  valid only for  $t_{a} > 0$ means 
that numerically the following must hold (in appropriate dimensionless units) i.e.
$$[ln T_{a}] [ln T_{b}] \leq {\pi\over 9\sqrt {3} m_{\mathrm Pl}}\eqno(43)$$
Therefore, the formalism described here has given a new way of describing fluctuations
of the background temperature. This estimate can be made as long as one is in the temporal
region where the lagrangian $(16)$ is valid and the conditions embodied in equations $(40)-(43)$ hold true. 
Note that there exist functional
time dependent parameters in the quantum amplitude, {\it viz.},
$\dot\nu(t_{a}),\dot\nu(t _{b}), \zeta(t_{b}, t_{a})$ which are related to solutions of the
Ermakov-Pinney equation. Now, the bounds on these functions may  be estimated from
phenomenological observations of satellite data if the fluctuations of the background
temperature can ever be accurately determined.

{\bf 5.Conclusion}

A unified formalism has been set up with a  $k-$essence lagrangian containing 
canonical kinetic  terms which provides a framework for estimating quantum fluctuations 
of the temperature in certain epochs. In an earlier epoch in the universe defined by
the scale factor being very small compared to the present epoch but very large compared
to the big bang and inflationary epoch (so that one is already in an expanding and flat universe),
this lagrangian is similar to an oscillator with time dependent frequency.
A sinusoidal perturbation from an ideal oscillator frequency leads to growing classical solutions 
for the scale factor and can be made consistent with the scenario of chaotic inflation as well 
as  with a radiation dominated universe at a certain 
epoch by suitable choice of parameters. As the time dependent frequency is basically related to 
a scalar field, it can also be taken to be similar to the "inflaton".Although (as stated in the 
beginning) we are strictly away from the inflationary scenario, we would like to have the scalar 
field to be as far as possible similar to that needed for inflation. This does away with the 
need for multiple scalar fields to account for specific features of  evolution of the universe. 
What has been shown here is that if the "inflaton" is given  an opportunity to influence the 
evolution for longer times, it can also help in contributing to the  dark matter and dark 
energy scenario.

Quantum fluctuations of the log of inverse temperature  can be estimated
in principle for a certain physical domain characterised by the epochs and the respective temperatures satisfying
certain bounds. In this region the expressions for the quantum fluctuations are valid.
At much later times, dark matter and dark energy can be be accounted
for as shown in \cite{gangopadhyay}.Establishing a formalism for  quantum fluctuations of 
the (inverse) temperature has been possible in this model owing to the presence of 
standard (canonical) kinetic terms in the lagrangian so that the usual prescriptions  
are applicable without ambiguity. This is the first model of this type. 

{\bf 5.Acknowledgments}

The author thanks the referee for extremely illuminating suggestions for improving
the manuscript.

The author also thanks the Centre For Astroparticle Physics and Space Science,
Bose Institute, Kolkata, for a sabbatical tenure during which this work was done and 
Sandip Chakrabarty for illuminating discussions.

\includegraphics[angle=-90,width=15cm]{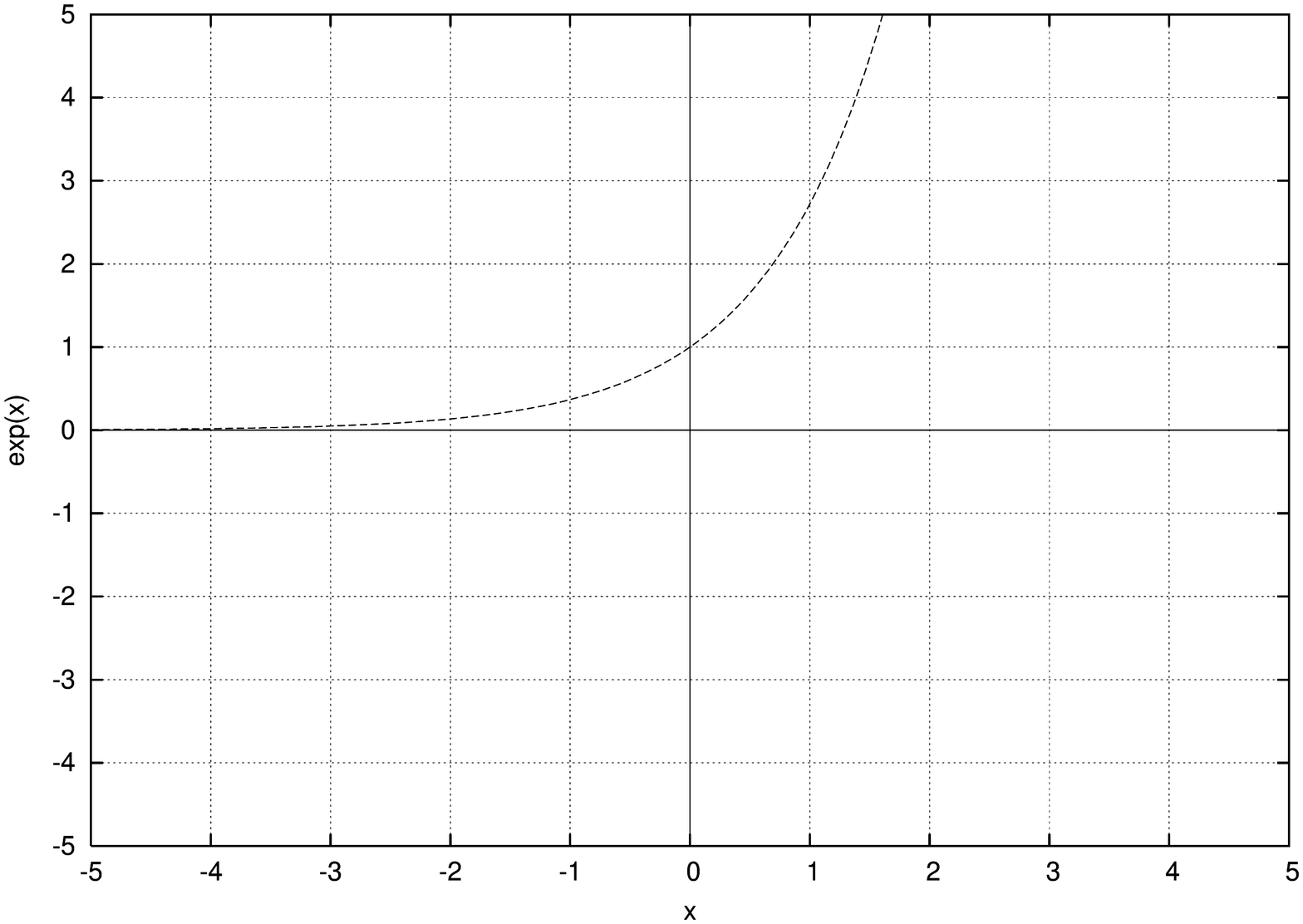}

\end{document}